%% file: paper.tex
\documentclass[11pt, final]{article}

\AtBeginDocument{%
  \providecommand\BibTeX{{%
    \normalfont B\kern-0.5em{\scshape i\kern-0.25em b}\kern-0.8em\TeX}}}

\usepackage{t1enc}
\usepackage[utf8]{inputenc}
\usepackage{authblk}
\usepackage{amsmath}
\usepackage{tikz} 
\usepackage[]{algorithm2e} 
\usepackage{listings} 
\usetikzlibrary{decorations.pathreplacing}
\usetikzlibrary{patterns} 
\usepackage{url}
\input{customlibs.tex}

\begin{document}


\date{}
\author[1,2]{Georges-Axel Jaloyan}
\author[3]{Konstantinos Markantonakis}
\author[3]{Raja Naeem Akram}
\author[1]{David Robin}
\author[3]{Keith Mayes}
\author[1]{David Naccache}
\affil[1]{DIENS, \'Ecole normale sup\'erieure, CNRS, PSL University, Paris, France}
\affil[2]{CEA, DAM, DIF, F-91297 Arpajon, France}
\affil[3]{Royal Holloway, University of London, London, United-Kingdom}

\title{Return-Oriented Programming on RISC-V}

\maketitle

\thanks{Modified version of our paper that appeared at AsiaCCS 2020.}

\begin{abstract}
	This paper provides the first analysis on the feasibility of Return-Oriented Programming (ROP) on RISC-V, a new instruction set architecture targeting embedded systems. We show the existence of a new class of gadgets, using several Linear Code Sequences And Jumps (LCSAJ), undetected by current Galileo-based ROP gadget searching tools. 
	
	We argue that this class of gadgets is rich enough on RISC-V to mount complex ROP attacks, bypassing traditional mitigation like DEP, ASLR, stack canaries, G-Free, as well as some compiler-based backward-edge CFI, by jumping over any guard inserted by a compiler to protect indirect jump instructions. 
	
	We provide examples of such gadgets, as well as a proof-of-concept ROP chain, using C code injection to leverage a privilege escalation attack on two standard Linux operating systems. Additionally, we discuss some of the required mitigations to prevent such attacks and provide a new ROP gadget finder algorithm that handles this new class of gadgets.
\end{abstract}

\section{Introduction}

Memory corruption vulnerabilities are one of the most popular entry points for hackers to hijack a program. Amongst them, stack overflow attacks have been popular since 1996 \cite{SmashingStack}. It was for long thought that the hacker would always inject some standalone payload, that could be detected as malicious as such, using methods such as executable space protection~\cite{DEP}. This assumption has been invalidated by \emph{Return-Oriented Programming} (ROP), introduced on par with the Galileo detection algorithm by Shacham in 2007~\cite{Shacham}, proving, as formulated by Dino Dai Zovi in 2010, that ``\textit{preventing the introduction of malicious code is not enough to prevent the execution of malicious computations}''~\cite{DinoDaiZovi}.

Since then, many countermeasures have been developed against ROP attacks~\cite{DROP,ROPDefender,GFree,PappasROP}. Each time, the publication of new ROP variants, such as JOP, SROP, SOP, or even JIT-spray~\cite{JOP2011,SROP,SOP2013,JITSPRAY} bypassed those stopgap mitigations. At the same time, these attacks have been extended to many architectures, including much simpler \emph{Reduced Instruction Set Computer} (RISC)~\cite{ROPSPARC}, confirming that those design flaws are widespread among all architectures. State-of-the-art mitigation methods such as \texttt{gcc}'s \texttt{-mmitigate-rop} option or G-Free~\cite{GFree}, tend to uproot such attacks by detecting and eliminating any code section that could be reused by an attacker, in the hope that the remaining gadgets would not be sufficient to mount complex attacks. Other even more radical methods like \emph{Control-Flow Integrity} (CFI) try preventing arbitrary control-flow transfers by validating the target of indirect jumps~\cite{Eurosys2010,CFIAbadi,ShadowStack}, often at the cost of performance, thus reducing their usability~\cite{CFIGross,CFIPayer}.

Likewise, these methods do hardly more than increasing the cost of such attacks, as it may be sufficient to find new unexpected gadgets to get back to step one of stack overflow exploitation. In this paper, we show once again, how to challenge the existing security mechanisms using a new class of gadgets that are undetected by the vast majority of published methods, based on the well-known Galileo algorithm. We explain how to produce such gadgets in RISC-V~\cite{RISCV}, a new \emph{Instruction Set Architecture} (ISA) which development began in 2010. Consequently, an attacker may be able to insert such gadgets in an open source program and exploit them unnoticed. 

RISC-V is based on the concept of RISC~\cite{RISC1}, targeting simplicity by providing few and limited computer instructions. RISC ISAs have become increasingly popular with the wide adoption of embedded devices such as smartphones, tablets, or other Internet of Things devices. The most popular RISC ISAs are currently ARM~\cite{ARMv8A}, Atmel AVR~\cite{AtmelAVR}, MIPS~\cite{MIPS}, Power~\cite{Power}, and SPARC~\cite{SPARC}.

RISC-V is the fifth RISC ISA published by UC Berkeley. It is completely free and open-source, with its User-Level ISA published in May 2017 in version 2.2. It features 32-bit and 64-bit little-endian variants (designated as \texttt{RV32} and \texttt{RV64}), with a future extension to 128 bits. While only expensive test boards feature RISC-V processors currently, many companies including Western Digital or Nvidia have announced the use of RISC-V chips in their future products~\cite{NVIDIA}. Hence, this architecture is of particular interest for such attacks, as most programs are in the process of being ported to this architecture, leaving the insertion of backdoors easy for an ill-intentioned programmer. 

We summarize our contributions as follows.
\begin{enumerate}
    \item We provide the first analysis on the feasibility of ROP attacks on the new RISC-V architecture.
    \item We introduce a new and stealthy class of ROP gadgets, undetected by all previously published methods based on the Galileo algorithm.
    \item We show the achievability of complex ROP attacks using this class of gadgets on RISC-V ISA, under the assumption of malicious C source code insertion generating such gadgets.
    \item We implement a proof-of-concept backdoored SUID program allowing privilege escalation on two standard Linux operating systems running on RISC-V, with every available ROP mitigation mechanism enabled.
    \item We present a new algorithm able to find ROP gadgets of this class and discuss the plausibility of their presence in existing RISC-V binaries.
\end{enumerate}

\section{Background}
In this section, we briefly introduce the key concepts related to this paper's scope-of-work and contributions. More particularly, we describe the memory corruption exploitation technique known as Return-Oriented Programming and detail some RISC-V features, later used in the paper.

\subsection{Return-Oriented Programming}
\label{sec:rop}
\begin{figure}[hb!]
	\centering
	\input{ropfig.tex}

	\caption{General principle of Return-Oriented Programming attacks. The vulnerability shown here consists in a buffer overflow from an unchecked \texttt{strcpy} allowing the user to smash the contents of the stack.}
    \label{fig:ropfig}
\end{figure}
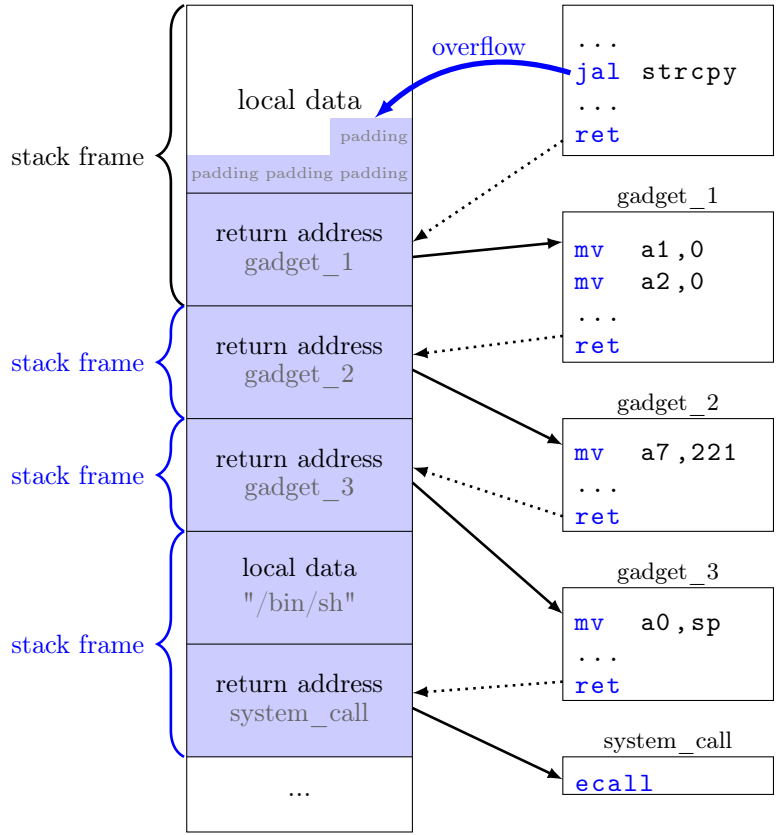

The first methods aiming at exploiting memory corruption bugs were as simple as a straightforward data injection into the program, which would end up being executed by the processor~\cite{SmashingStack}. The introduction of \emph{Data Execution Prevention} (DEP)~\cite{DEP} made those attacks almost impossible, as injected data could not be executed anymore. In this battle between the shield and the sword, \emph{Return-Oriented Programming} (ROP) has been the answer from malware developers. The first ROP attack was publicly presented in 2001 by Nergal in Phrack~\cite{PhrackReturnLibC}. 

As shown by Fig.~\ref{fig:ropfig}, it bypasses DEP by injecting in the stack a succession of call frames. Each call frame will result in the execution of a \emph{gadget}: a small snippet of legitimate code containing a small number of instructions ended by a \texttt{ret}. When the \texttt{ret} instruction is reached, the address of the next gadget is popped from the stack into the program counter. Provided that enough different gadgets are available in the executable, arbitrary code may be executed by chaining those gadgets.

Two categories of gadgets can be distinguished. The first one using only legitimate code written by the programmer, also called the \emph{Main Execution Path} (MEP). The second category uses overlapping code, called \emph{Hidden Execution Path} (HEP), \emph{i.e.} code sections that have another interpretation by the CPU depending on its internal status (32 or 64 bits, Thumb mode, or on the offset at which the execution has started). The latter has the advantage of bypassing any compiler-added stack protection mechanism, presenting a wider variety of side-effects and being undetectable by traditional linear or recursive disassemblers, which only handle the MEP of a program. 
	
The first academic paper studying this technique was published in 2007 by Shacham~\cite{Shacham}, in which he presents ROP on x86 and the Galileo algorithm, which allows the detection of gadgets in any executable memory region. It is based on a backward disassembly method, starting from every return instruction, and then trying to recursively bruteforce the length of the previous instruction. This provides a tree of possible gadgets all ending with a return. 

The most common attack scheme consists in scanning the executable sections of the binary with Galileo based~\cite{XROP,IDAsploiter,IRGadget} or with other \textit{ad hoc} algorithms~\cite{ROPgadget} to find gadgets which are thereupon used to devise a ROP chain performing the required computation. Intermediate languages are sometimes used, allowing the design of higher-level ROP chains that are then compiled to the gadget language~\cite{Angrop,IRGadget}. Finally, the payload is adapted to the injection method, with techniques like padding, \texttt{NUL} bytes removal, or even alphanumeric conversion, which are not within the scope of this study.

By design, the Galileo algorithm is only able to find gadgets made of a straight-line instruction sequence, with no jumps except for the last instruction. Such a sequence is called a \emph{Linear Code Sequence And Jump} (LCSAJ). Gadgets spanning over several LCSAJs are thus undetected by Galileo, and, to the best of our knowledge, have never been subject to study in the context of ROP attacks.

\subsection{RISC-V}

RISC-V splits its instruction set between a mandatory core set (\texttt{RV64I}) and different optional extensions, each of which is designated by a letter. The defined extensions include integer multiplication and division~(\texttt{M}), atomic operations~(\texttt{A}), single-, double- or quad-precision~(\texttt{F}, \texttt{D}, \texttt{Q}) floating-point operations, decimal floating-point operations~(\texttt{L}), compressed instructions~(\texttt{C}), bit-manipulation~(\texttt{B}), just-in-time~(\texttt{J}), transactional memory~(\texttt{T}), packed-SIMD instructions~(\texttt{P}), vector operations~(\texttt{V}), and user-level interrupts~(\texttt{N}). The general purpose ISA, which includes \texttt{IMAFD}, is designated by the letter \texttt{G}. In what follows, we focus on the \texttt{RV64GC} ISA, which is the one agreed on by Debian and Fedora porters, as well as members of the RISC-V Foundation. On top of that, the Foundation intends to provide ``\textit{a profile for standard RISC-V Unix platforms that will include \texttt{C} extension as mandatory}''.\footnote{\url{https://wiki.debian.org/RISC-V}}
	
There are 31 general purpose 64-bit registers (named \texttt{x1}-\texttt{x31}), 32 floating-point registers (\texttt{f0}-\texttt{f31}), a program counter (\texttt{pc}), as well as various control-and-status registers. The pseudo-register \texttt{x0} designates the zero constant.
RISC-V provides a standard ELF \emph{Application Binary Interface} (ABI), called psABI~\cite{RVELFABI}, with the naming convention provided in Fig.~\ref{Fig:psABI}. 
	
	\begin{figure}[h]
		\centering
		\begin{tabular}{|l|c|l|}
			\hline
			Register & ABI Mnemonic & Meaning \\
			\hline
			\texttt{x0} & \texttt{zero} & Zero \\
			\texttt{x1} & \texttt{ra} & Return address \\
			\texttt{x2} & \texttt{sp} & Stack pointer \\
			\texttt{x3} & \texttt{gp} & Global pointer \\
			\texttt{x4} & \texttt{tp} & Thread pointer \\
			\texttt{x5}-\texttt{x7} & \texttt{t0}-\texttt{t2} & Temporary registers \\
			\texttt{x8}-\texttt{x9} & \texttt{s0}-\texttt{s1} & Callee-saved registers \\
			\texttt{x10}-\texttt{x17} & \texttt{a0}-\texttt{a7} & Argument registers \\
			\texttt{x18}-\texttt{x27} & \texttt{s2}-\texttt{s11} & Callee-saved registers \\
			\texttt{x28}-\texttt{x31} & \texttt{t3-}\texttt{t6} & Temporary registers \\			
			\hline
			\hline
			\texttt{f0}-\texttt{f7} & \texttt{ft0}-\texttt{ft7} & Temporary registers \\
			\texttt{f8}-\texttt{f9} & \texttt{fs0}-\texttt{fs1} & Callee-saved registers \\
			\texttt{f10}-\texttt{f17} & \texttt{fa0}-\texttt{fa7} & Argument registers \\
			\texttt{f18}-\texttt{f27} & \texttt{fs2}-\texttt{fs11} & Callee-saved registers \\
			\texttt{f28}-\texttt{f31} & \texttt{ft8}-\texttt{ft11} & Temporary registers \\
			\hline
			
		\end{tabular}
		\caption{Naming convention for registers, per RISC-V ELF psABI.}
		\label{Fig:psABI}
	\end{figure}

While most RISC ISAs require naturally aligned instructions, \texttt{RV64GC} features 32-bit and 16-bit instructions, aligned on 16 bits, like in Thumb-2 extension introduced with ARMv6T2~\cite{ARMv6T2}. Instruction length is encoded in the least-significant byte (hence with the lowest address as RISC-V is little-endian): 16-bit instructions require the last two bits to be different from \texttt{11} whereas 32-bit instructions have their last two bits equal to \texttt{11} with the three previous bits different from \texttt{111}. 

\begin{figure}[h!]
\vspace*{-0.4cm}
	\centering

\begin{tikzpicture}[scale=0.7, every node/.style={scale=.8}]
\draw(0,0) rectangle (12,1);
\draw(2,0) to (2,1);
\draw(4,0) to (4,1);
\draw(6,0) to (6,1);
\draw(8,0) to (8,1);
\draw(10,0) to (10,1);

\node[text centered, above] at (1,1) {\textbf{\texttt{13}}};
\node[text centered, above] at (3,1) {\texttt{\textbf{4f}}};
\node[text centered, above] at (5,1) {\texttt{\textbf{83}}};
\node[text centered, above] at (7,1) {\texttt{\textbf{23}}};

\node[text centered, below] at (5,0) {\texttt{\textbf{83}}};
\node[text centered, below] at (7,0) {\texttt{\textbf{23}}};
\node[text centered, below] at (9,0) {\texttt{\textbf{0b}}};
\node[text centered, below] at (11,0) {\texttt{\textbf{00}}};

\node[text centered] at (1,0.5) {\texttt{000\textcolor{blue}{\textbf{10011}}}};
\node[text centered] at (3,0.5) {\texttt{01001111}};
\node[text centered] at (5,0.5) {\texttt{100\textcolor{blue}{\textbf{00011}}}};
\node[text centered] at (7,0.5) {\texttt{00100011}};
\node[text centered] at (9,0.5) {\texttt{00001011}};
\node[text centered] at (11,0.5) {\texttt{00000000}};

\draw[decoration={brace,raise=0pt, amplitude=10pt},decorate,line width=1pt,black] 
(0,1.3) -- (8,1.3) node [black, midway, xshift=0cm, yshift=0.7cm] {\Large \texttt{\textcolor{blue}{xori} t5,t1,568}} ;

\draw[decoration={brace,raise=0pt, mirror,amplitude=10pt},decorate,line width=1pt,black] 
(4,-0.3) -- (12,-0.3) node [black, midway, xshift=0cm, yshift=-0.7cm] {\Large \texttt{\textcolor{blue}{lw} t2,0(s6)}} ;
\end{tikzpicture}

\vspace*{-0.2cm}
\caption{Two 32-bit overlapping instructions of $\boldsymbol{\mathrm{I}_{1}}$ (little-endian representation). Instruction length encoding for each instruction is emphasized in blue.}
\label{fig:instructions1}
\end{figure}
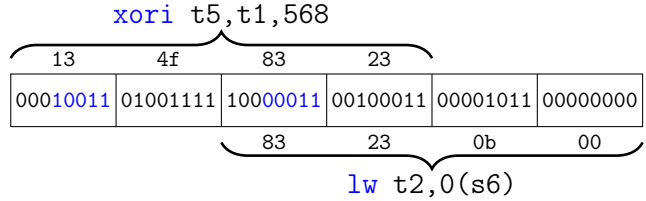

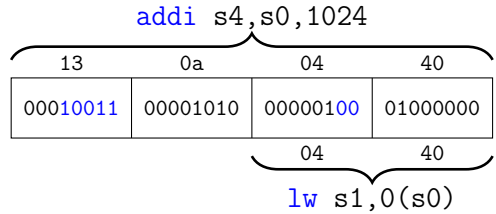
\begin{figure}[h!]
	\centering
	
	\begin{tikzpicture}[scale=0.8, every node/.style={scale=0.8}]
	\draw(0,0) rectangle (8,1);
	\draw(2,0) to (2,1);
	\draw(4,0) to (4,1);
	\draw(6,0) to (6,1);

	\node[text centered, above] at (1,1) {\textbf{\texttt{13}}};
	\node[text centered, above] at (3,1) {\texttt{\textbf{0a}}};
	\node[text centered, above] at (5,1) {\texttt{\textbf{04}}};
	\node[text centered, above] at (7,1) {\texttt{\textbf{40}}};
	
	\node[text centered, below] at (5,0) {\texttt{\textbf{04}}};
	\node[text centered, below] at (7,0) {\texttt{\textbf{40}}};
	
	\node[text centered] at (1,0.5) {\texttt{000\textcolor{blue}{\textbf{10011}}}};
	\node[text centered] at (3,0.5) {\texttt{00001010}};
	\node[text centered] at (5,0.5) {\texttt{000001\textcolor{blue}{\textbf{00}}}};
	\node[text centered] at (7,0.5) {\texttt{01000000}};
	
	\draw[decoration={brace,raise=0pt, amplitude=10pt},decorate,line width=1pt,black] 
	(0,1.3) -- (8,1.3) node [black, midway, xshift=0cm, yshift=0.7cm] {\Large \texttt{\textcolor{blue}{addi} s4,s0,1024}} ;
	
	\draw[decoration={brace,raise=0pt, mirror,amplitude=10pt},decorate,line width=1pt,black] 
	(4,-0.3) -- (8,-0.3) node [black, midway, xshift=0cm, yshift=-0.7cm] {\Large \texttt{\textcolor{blue}{lw} s1,0(s0)}} ;
	\end{tikzpicture}
\vspace*{-0.2cm}
	\caption{A 32-bit instruction of $\boldsymbol{\mathrm{I}_{2}}$ whose last 2 bytes are also a 16-bit valid instruction (little-endian representation)}
	\label{fig:instructions2}
\end{figure}

Combining those two peculiarities of \texttt{RV64GC} opens the door to overlapping instructions, that can be obtained by either using two 32-bit instructions 2 bytes apart (Fig.~\ref{fig:instructions1}), or by using a 32-bit instruction whose last 2 bytes are also a valid 16-bit compressed instruction (Fig.~\ref{fig:instructions2}). In what follows, we use $\mathrm{I}_{1}$ to designate the set of 32-bits instructions allowing overlapping sequences, whereas the set of 32-bit instructions whose last 2 bytes are valid 16-bit instruction will be denoted by $\mathrm{I}_{2}$. Examples of overlapping for both sets $\mathrm{I}_{1}$ and $\mathrm{I}_{2}$ are given in Fig.~\ref{fig:instructions1} and \ref{fig:instructions2}. Typically, an overlapping sequence consists of several instructions of $\mathrm{I}_{1}$ chained together, optionally ending with an instruction of $\mathrm{I}_{2}$. 

\section{Threat model and attack overview}
In this section, we explicit our target platforms, aiming run-of-the-mill RISC-V systems with off-the-shelf ROP mitigations deployed. We also present two attack scenarios taking advantage of our new class of gadgets for improved concealment.

Our target platform features a standard Linux operating system, such as Debian or Fedora, with two levels of privilege, that we call user and root. Standard protections are deployed, such as \emph{Address Space Layout Randomization} and \emph{Data Execution Prevention}, that prevent common stack overflow exploits. Programs are compiled with the standard \texttt{gcc} provided by the operating system, adding \texttt{gcc}'s ROP mitigation mechanism using compiler flag \texttt{-fstack-protector-strong}. Note that some other mitigation specific to x86 are not available on RISC-V, like \texttt{gcc}'s \texttt{-mmitigate-rop} option or \texttt{clang}'s CFI. In Section~\ref{sec:mitigations}, we discuss the ability of such mitigations, if ported to RISC-V, to hamper attacks using this new class of gadgets. 
\subsection{Closing (stealthily) the gap between vulnerability and exploitation}

The first attack scenario focuses on adding a backdoor to a program leading to a ROP attack. Backdoors allow any person aware of their existence to reach a \emph{privileged state} upon a specific \emph{input}. In order to create a backdoor, two distinct elements must be stealthily inserted by an attacker: a \emph{trigger} and a \emph{payload}~\cite{Backdoors}. In our scenario, we assume the attacker already managed to insert a trigger (or find an existing one), in the form of a \emph{ROP exec vulnerability}: a one-time memory write like a buffer overflow combined with an arbitrary control-flow redirect, such as a return at the end of the function, use-after-free, type confusion, or even corrupted instruction through fault injection~\cite{FAULT2016}. Such vulnerabilities are pretty common in programs, and are often rendered non exploitable by reducing the number of available gadgets and by deploying ROP mitigations, such as ASLR, stack canaries, backward-edge CFI, or G-Free.

To lower the bar of exploitability, the attacker must embed gadgets in the payload of his backdoor, aiming at preventing any unaware outsider from stumbling upon those gadgets. As a steppingstone for future elaboration, we consider generic C code injection through traditional backdooring, as we believe that one variant of this scenario may target C++ Just-in-Time compilers (like Cling~\cite{Cling} or ClangJIT~\cite{ClangJIT}, once they get ported to RISC-V) to mount JIT-spraying attacks~\cite{JITSPRAY2018}. Indeed, identical assumptions are required for the latter: code injection and ROP exec vulnerability.

As an illustration, we consider the case where the attacker has a user privileged level access to the system, including shell, ability to run programs, read access to binaries and libraries. The goal of the attacker is to increase his privilege level to root, which in practice thoroughly compromises the system by granting a read-write access to the whole target. Such an attack is called a \emph{privilege escalation attack}, and is at the core of highly publicized attacks such as iOS jailbreaking~\cite{iOSjailbreak}. To that end, the attacker will use a program that can be executed by the user, but running at a root privilege level. Those programs are called SUID (\emph{Set owner User ID up on execution}), and are abundant on any system. Indeed, actions as simple as changing a password, plugging a USB key or granting root privilege for an authorized user require the execution of SUID programs. 

To backdoor such programs, the attacker may upstream underhanded C code in some open-source project. Details on how to achieve this have been provided by Gilbertson~\cite{Hackernoon} and thoroughly studied by Prati~\cite{Prati}, with some examples provided in the \emph{Underhanded C Contest}\footnote{http://www.underhanded-c.org/} and DEF CON's \emph{Hiding backdoor in plain sight} contest. Here, the payload consists in a set of ROP gadgets that span over several LCSAJs. Furthermore, those gadgets are using overlapping techniques, so that only the last LCSAJ is in the MEP, whereas all the previous ones are in the HEP, thus hiding them to currently available ROP gadget searchers. To trigger the exploit and gain root access, the attacker only has to execute the SUID program with the adequate user input. 

\subsection{Creating a (concealed) persistent backdoor on a compromised system}

The second attack scenario leverages privilege escalation through SUID to build a persistent backdoor in a compromised target. Persistence is considered as a key step in a complex attack chain to maintain access into compromised systems upon slight environment changes (reboot, updates, password change). This attack is much easier to implement than inserting backdoors in highly scrutinized SUID programs, as it requires the attacker to only get a one-time root access, and grant SUID permission to a program for which he has knowledge of the existence of a privilege escalation exploit. Such backdoors are quite common,\footnote{https://attack.mitre.org/techniques/T1166/} as they involve modifying the permissions of only one file, which is not monitored by default on popular intrusion detection systems such as \texttt{rkhunter}, \texttt{chkrootkit}, or \texttt{samhain}.

For better chances of success, this can be combined with the first attack scenario, by inserting hidden gadgets in a non SUID open-source program, which is much easier to achieve. This backdoored program, embedding the hidden gadgets and a ROP exec vulnerability, will be legitimately deployed on the target. Should a security analyst audit the program before the attack, he will wrongly conclude that the vulnerability is not exploitable, hence not requiring an urgent patch. After the attack has been discovered, even if a forensics analyst comes across the program granted with SUID permission, without the knowledge of the ROP-chain, he will waste precious time and effort trying in vain to identify the mechanism allowing privilege escalation.

\begin{figure*}[ht!]
\centering
\input{gadget.tex}
\caption{Segmentation of the different code sequences present in \texttt{function15c}. The gadget is highlighted in gray.}
\label{fig:function1}
\end{figure*}

\section{Inserting Hidden Gadgets}

\label{sec:backdoor}
For the sake of realism, we intend to use code created by a standard C compiler like \texttt{gcc}. We create exactly one function per gadget (named \texttt{function1}, ...), each ending with a C \texttt{return} instruction. For each function, the compiler may add assembly code at the beginning and the end of the function whose purpose is to respectively insert (\emph{save} sequence) and remove (\emph{restore} sequence) the call frame from the stack, depending on whether a callee-saved register is modified by the function. Inserting a nested call in the function is an easy way to be sure that the compiler will emit these save and restore sequences.

Indeed, the presence of a restore sequence is crucial for mounting a ROP attack, as we need to tamper with the return address register  \texttt{ra}, which is callee-saved. Inserting malicious call frames into the stack hence grants control over the program counter through \texttt{ra}. In practice, a vast majority of functions do call other functions, either in the program, or in any library. In our proof-of-concept attack, we purposely added a call to a dummy function in every gadget function. Other ROP variants using alternative control-flow instructions such as indirect jumps or exceptions are beyond the scope of this paper.

The malicious gadget is made of two LCSAJs, the first being hidden with code overlapping and the last being the legitimate restore sequence. A detailed example for one of the gadgets is provided in Fig.~\ref{fig:function1}. The C code (using \texttt{gcc -Os -fstack-protector-strong}) used to generate it is:
\begin{lstlisting}[language=C]
long long function15c(){
    dummy();
    dummy4((signed) 0x9932000, 
           0, 
           (signed) 0xa0212000,
           (signed) 0x23371000);
    return 0;}
\end{lstlisting}

The hidden instructions are directly written in C code, and feature one or two instructions followed by a jump to a relative offset. In the example of Fig.~\ref{fig:function1}, the MEP consists of two 32-bit $\mathrm{I}_{1}$ instructions followed by one $\mathrm{I}_{2}$ instruction, whereas the HEP comprises two 32-bit $\mathrm{I}_{1}$ instruction followed by one 16-bit jump instruction. Here, the jump is only 8 bytes off its target, but it is definitely possible to modify this value to hide the overlapping LCSAJ anywhere, even in other functions. In this gadget, magic constants are loaded into the arguments of a function. The other gadgets use a mix of arithmetical and floating-point operations, as well as load and stores instructions. To have a consistent output among different compiler versions and environments, we forced register allocation (using the \texttt{register} keyword), and prevented instruction reordering in the overlapping sequence. Magic constants as arguments of the function cannot be prevented, as the opcode of a HEP instruction lies in the operand of the MEP instruction. However, many source code obfuscation techniques may come to help here, such as C-preprocessor~\cite{CPREPROCESSOR} or lightweight constant blinding, hiding the magic constants respectively until the preprocessing and constant folding passes of the compiler.

\section{Chaining the Gadgets}

\label{sec:chain}
\begin{figure}
\begin{lstlisting}[xleftmargin=0pt,basicstyle=\small\ttfamily]
8    slti  t2,t2,225
24   slti  t2,t2,225    //t2:=1
40   slti  t2,t2,225    //NOP
48   .plt_address+1823
56   slti  a1,t2,-1999  //a1:=0
72   mul   a4,t2,sp     //a4:=.base+80
88   slti  t2,t2,-1999  //t2:=0
104  slti  a2,t2,-1999  //a2:=0
120  addi  a4,a4,-1278
136  addi  a4,a4,1275   //a4:=.base+77
152  addi  t2,t2,-31    //t2:=-31
168  ld    s6,-29(a4)   //s6:=.plt+1823
184  ld    s6,-1823(s6) //s6:=.__libc_start_main@libc
200  addi  t1,s6,-1823
208  .ecall1_offset+1823
216  addi  s11,t1,s2    //s11:=.setuid@libc:34
232  sd    s11,315(a4)  //.base+392<-s11
248  addi  s3,a4,363    //s3:=.base+440
264  sd    s3,307(a4)   //.base+384<-.base+440
280  sd    s3,363(a4)   //.base+440<-.base+440
296  addi  t1,s6,-1823
304  .ecall2_offset+1823
312  addi  s11,t1,s2    //s11:=.setuid@libc:38
328  sd    s11,411(a4)  //.base+488<-s11
344  addi  t2,t2,-31    //t2:=-62
360  addi  t2,t2,-31    //t2:=-93
376  sltiu a0,t2,2017   //a0:=0
384  0                  //.base+440
392  0  //ecall1 at .setuid@libc:34
440  0                  // stack canary
456  addi  a7,t2,314    //a7:=221
472  addi  a0,a4,67     //a0:=.base+507
488  0  //ecall2 at .setuid@libc:38
507  "/bin/sh"\end{lstlisting}
\caption{High-level description of the ROP chain. The first column describes the offset in bytes relative to the beginning of the ROP chain. The notation with a leading dot \texttt{.xxx@yyy:off} designates the address of \texttt{xxx} in \texttt{yyy} at offset \texttt{off}. The notation \texttt{<-} designates a memory store, and \texttt{:=} an assignment. The \texttt{.ecall1\_offset+1823} indicates the location where we put the offset of the \texttt{ecall} instruction in the \texttt{setuid} function of the C library relative to the \texttt{\_\_libc\_start\_main} function. Similarly, the \texttt{.plt\_address+1823} indicates the location where the PLT address should be inserted. }
    \label{fig:asm}
\end{figure}

In the previous section, we described our method to build one gadget hiding some $\mathrm{I}_{1}$ instructions. In our full privilege escalation attack, we need to chain several of such gadgets together. We will aim at spawning a root shell, by invoking two system calls, the first being \texttt{setuid(0)} and the second \texttt{execve("/bin/sh",0,0)}. 

In RISC-V, each syscall requires the execution of a special instruction named \texttt{ecall}, with register \texttt{a7} set to a value encoding the call.\footnote{\url{https://www.lurklurk.org/syscalls.html}} For each call, one or several arguments may be passed, in registers \texttt{a0}, \texttt{a1}, \texttt{a2}, ... The \texttt{setuid} syscall requires \texttt{a7} to be set to 146, and \texttt{a0} to the desired userid, in our case zero. The \texttt{execve} syscall requires register \texttt{a7} to be set to 221 (\texttt{0xdd}), \texttt{a1} and \texttt{a2} to zero, and \texttt{a0} to point to the address of the string \texttt{/bin/sh}. The next paragraphs explain how to achieve this result by using only $\mathrm{I}_{1}$ instructions. We summarize the high-level overview of the ROP chain in an assembly-like pseudocode in Fig.~\ref{fig:asm}. The link to the full source code is available in Appendix~\ref{app:code}.

Let us start by zeroing (resetting) a register. For this purpose, we use the \texttt{slti} instruction (store less than immediate), that compares its source register to a constant, and if lower resets the destination register, else sets it to 1. By performing two \texttt{slti} instructions with a negative immediate and with same source and destination register, we are guaranteed to reset the register. In Fig.~\ref{fig:asm}, this happens at offset 88. We can then reset other registers by just performing an \texttt{slti} with a zeroed source register and a negative immediate (offset 104).  

The execution of an \texttt{ecall} instruction is trickier, as $\texttt{ecall} \not\in \mathrm{I}_{1,2}$. Hence, we must find an existing \texttt{ecall} and insert its location into the stack, so that the program counter points to it after the execution of the last gadget. If the program is statically compiled, this does not raise any issue. However, in most operating systems, the program is compiled dynamically, which results in every \texttt{ecall} instructions to be located in the libraries. Consequently, in order to find the address of such an instruction, we must outsmart the \emph{Address Space Layout Randomization} (ASLR), which loads the linked libraries at random addresses. Randomized libraries are then linked to the program through the \emph{Procedure Linkage Table} (PLT), in which the dynamic loader (\texttt{ld.so}) stores the randomized addresses of each external function called by the program. The PLT itself is always stored in the same memory area, statically known to er (offset 48). Programs compiled as \emph{Position Independent Executable} with \texttt{-fPIE} require an information leak to locate the PLT. By reading into the PLT, we compute the address of our \texttt{ecall} instruction and write it into the stack, so that the last gadget before the \texttt{ecall} will pop its address and jump on it, triggering the syscall. 

If a program uses the standard C library, then an initialization function called \texttt{\_\_libc\_start\_main} is systematically included in the PLT. In version 2.27 of the library, there is an \texttt{ecall} at offset 220, making a perfect candidate for the \texttt{execve} syscall. However, this instruction is not satisfactory enough for our \texttt{setuid} syscall, as we need to continue the execution of our ROP chain after invoking the syscall. Here, the candidate is part of an infinite loop.

One may think that jumping at the beginning of the \texttt{setuid@libc} function of the C standard library may be a good idea. This is definitely not, as the function inserts its own call frame into the stack, based on the value of \texttt{ra} at its entry. Since we already use \texttt{ra} to hijack the control flow with \texttt{ret} instructions, the function would return at its beginning, causing an infinite loop. Jump and link instructions that could modify \texttt{ra} are inadequate as well, inasmuch as they are detectable by Galileo.

Our solution involves jumping directly into the middle of the \texttt{setuid@libc} function, making use of the instruction that sets register \texttt{a7} to 146 immediately followed by the \texttt{ecall}. As a downside, we now must bypass \texttt{gcc}'s \emph{stack protector} (SP), that enforces backward-edge control-flow integrity, obliging the function to return to its caller. Concretely, it checks whether the call frames have been tampered with by generating a random number, the \emph{canary}, at the beginning of the function, and storing it in two different locations. During the restore sequence, the two values are compared, and, if different, the program aborts.

Howbeit, the other location at which the canary is stored is pointed to by \texttt{s0}, which happens to be a callee-saved register, also used by \texttt{gcc} as a frame pointer. Hence it may be possible to obtain a gadget whose restore sequence pops \texttt{s0} from the stack, which allows hijacking the canary. We do so by writing at offset 384, which smashes the value of \texttt{s0}, thence pointing both copies of the canary to the same memory area. In this way, the canary test will always pass, as both pointers are now aliased. Finally, the gadget at offset 232 inserts into the stack the address of the \texttt{ecall} in \texttt{setuid@libc} using the location of \texttt{\_\_libc\_start\_main} obtained through the PLT.

The \texttt{execve} syscall is easier to prepare. We reset \texttt{a2}, and straightforwardly set \texttt{a7} to 221. The gadget at offset 328 inserts into the stack the address of the \texttt{ecall} candidate, also in \texttt{setuid@libc}. Note that we do not need to bypass SP this time, as the \texttt{execve} syscall will spawn a new process. Finally, we take advantage of the previously leaked stack pointer (at offset 72) to set \texttt{a0} to the address of the string \texttt{/bin/sh}, located after the last call frame of our ROP chain. 

\section{Attack Proof-of-Concept on Different Platforms}
\label{sec:experimentation}

In this section, we experiment our attack on two Linux operating systems, Debian and Fedora, running as a chroot environment on a HiFive Unleashed development board, featuring a quad-core Freedom U540 \texttt{RV64GC} processor. 

\subsection{Debian chroot on HiFive Unleashed}

We first try our attack on the HiFive Unleashed board with a reduced Linux buildroot system shipped with the board. We add a Debian chroot, allowing the access of Debian features within the minimal operating system. Additionally, we create an unprivileged user, setting up the stage for our attack. Given that there is no \texttt{gcc} available on Debian RISC-V, we statically cross-compile the binary from another host computer. Static compilation greatly simplifies our attack, as all the libraries are now included within the program, rendering ASLR ineffective. Nevertheless, we still use \texttt{-fstack-protector-strong}, to harden the program against ROP attacks. 

Compared to previous scenario, we do not need to access the PLT anymore. Instead we need to find an \texttt{ecall} in the program itself. For this purpose, the function \texttt{\_\_internal\_atexit} is a perfect candidate. Indeed it is always included by default in binaries using the standard C library, and remarkably, falls through the cracks of SP. We write new gadgets in handwritten assembly this time, and adapt the ROP chain.

The test program embeds the gadgets, whose construction is detailed in Section~\ref{sec:backdoor}, and the ROP chain with some simplifications compared to Section~\ref{sec:chain}. Finally, a function with a ROP exec vulnerability is added to the program, whose sole purpose is to grant the attacker the possibility to smash the stack, launching the attack upon return. We use an assembly instruction that straightforwardly replaces the stack by the ROP chain, which produces similar results as a buffer overflow vulnerability that arises from a \texttt{scanf("\%s",buffer)}. 

After setting the SUID permission using \texttt{chmod u+s} to the binary, the user logs in and executes the target program, successfully spawning a root shell.

\subsection{Fedora}

\begin{figure}[!t]
    \centering
    \includegraphics[width=0.97\textwidth]{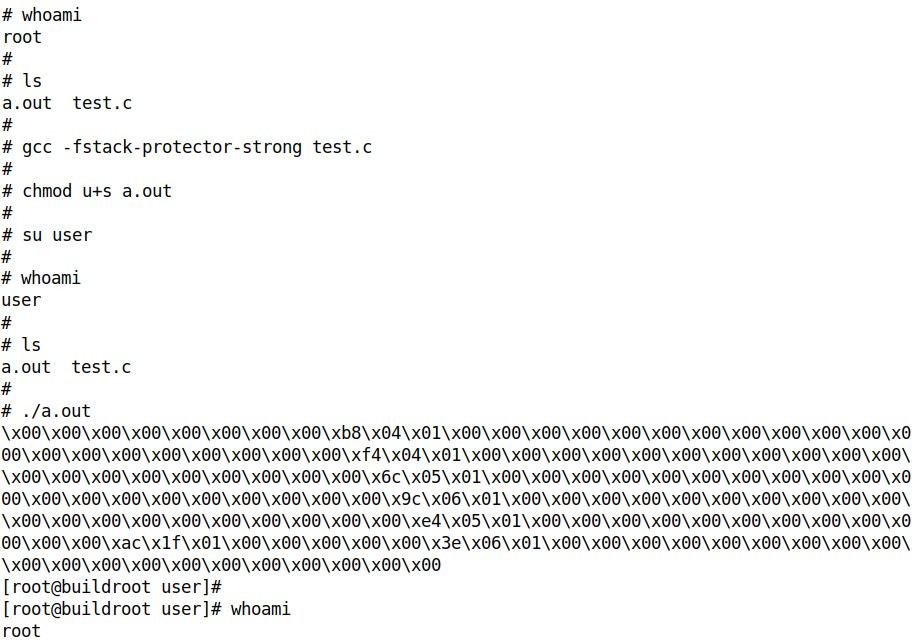}
    \caption{The attack setup with the Hifive Unleashed board featuring a Fedora chroot. A serial connection on the micro-usb port allows a user-level access to the board. An SUID executable in the user's home directory allows a successful privilege escalation attack, upon injecting the ROP chain (cropped). }
    \label{fig:hifiveu}
\end{figure}

We then moved to a Fedora 28 stage 4 disk image, another Linux based OS with many more features. It has a package manager with a \texttt{gcc} version 7.3.1 able to dynamically compile programs directly on the board with a standard C library in version 2.27.\footnote{\url{https://fedorapeople.org/groups/risc-v/disk-images/}} Our attack was successfully tested both on the RISC-V Fedora powered by a QEMU virtual machine~\cite{QEMU} and a Fedora chroot for Linux buildroot running on the HiFive Unleashed board, shown in Fig.~\ref{fig:hifiveu}. 

As we expected, we did not witness any difference between both tests, as QEMU emulates a HiFive Unleashed \texttt{RV64GC} board, without some of its micro-architectural features like caches or timings. Moreover, in both cases, ASLR is set to conservative randomization mode, which randomizes the stack, VDSO page, and shared memory region position. The binary itself is not randomized, which creates the opportunity of such code-reuse attacks. The data segment base is located immediately after the end of the executable code segment. We successfully bypass ASLR and SP, using the method presented in Section~\ref{sec:chain}. 

Likewise, our test program embeds the malicious gadgets written in C, the ROP chain and the ROP exec vulnerability. The program is compiled by root using the standard \texttt{gcc} with options \texttt{-Os -fstack-protector-strong}, and given SUID permission using \texttt{chmod u+s}. The user then logs in and executes the program, again successfully escalating privilege.

\section{Proposed Countermeasures}
\label{sec:mitigations}
In this section, we review different methods that could be implemented to reduce the threat posed by the new gadgets described in this paper, from the simplest to the most complex solutions. We also provide a new algorithm for finding gadgets in RISC-V, that aims at improving and replacing the Galileo algorithm in ROP gadget finders.

Although we managed to bypass \texttt{gcc}'s SP, we believe that stack canaries may still be useful, as they try to prevent stack smashing, reducing the number of ROP exec vulnerabilities, and partial function execution, reducing the number of MEP gadgets, thus raising the cost for ROP attacks. In our attack scenario, even if SP is deployed everywhere (using option \texttt{-fstack-protector-all}), our gadgets are still able to jump over any canary check directly on the restore sequence, rendering them ineffective. Therefore, we recommend checking the canary immediately before the return rather than at the beginning of the restore sequence, as done by various CFI implementations.

In \texttt{gcc}, stack canaries are deployed using three different compilation flags: \texttt{-fstack-protector-all} that adds stack canaries to every function (but not to glue-code), \texttt{-fstack-protector} for only the most vulnerable functions (calling \texttt{alloca}, or containing a buffer whose size is larger than 8 bytes), and \texttt{-fstack-protector-strong}, introduced in 2012 that strikes a balance in between. Since Fedora 20, all packages are compiled with the last option. Thus, compiling all SUID programs with option \texttt{-fstack-protector-all}, as done on FreeBSD, can prove to be a good mitigation, as it widens the gap between vulnerability and exploit by reducing the number of available gadgets. Thence, an attacker would need to embed more hidden gadgets in his payload, increasing the probability of being detected.

If we consider compiler-based backward-edge CFI variants like LLVM-CFI,\footnote{\url{https://clang.llvm.org/docs/ControlFlowIntegrityDesign.html}} MCFI or Picon~\cite{CFIPayer,MCFI,Picon}, the restore sequence may be hardened in a way that may not allow reusing any part of it, \emph{e.g.} by putting the target validation guard between the return and the assignment to \texttt{ra} from the stack. This leaves us with only the last return instruction that can be jumped to from the HEP. Although we hypothesize it may be possible to assign any value to \texttt{ra} directly from the HEP, it is actually much easier to fall back on the restore sequence of another function that is not protected by compiler-based CFI, like glue-code. For the C standard library, the \texttt{\_\_libc\_csu\_init} function of \texttt{crt1.o} inserted by \texttt{gcc} and \texttt{clang} is a perfect candidate, as it contains an unprotected restore sequence, even when compiled with SP (\texttt{-fstack-protector-all}) and LLVM-CFI (\texttt{-fsanitize=cfi} on clang). 

OpenBSD has its own SP version called \emph{RetGuard}~\cite{OpenBSD}, running on par with gadget reduction techniques, with the same shortcomings as \texttt{gcc}'s SP. More generally, gadget reduction techniques like G-Free~\cite{GFree} or code randomization~\cite{Randomization} intend to eliminate any unaligned indirect jump, relying on canaries or backward-edge CFI to prevent malicious use of aligned branches, which is effective only against gadgets having one LCSAJ. The new gadgets presented in this paper fall out of reach of those mitigations.

To include this new class of gadgets in existing mitigation, we would have combine them with a static analysis pass verifying that every main and hidden execution path ending with an indirect jump does go through the canary check (SP) or reaches target validation (backward-edge CFI). For this purpose, we provide Algorithm~\ref{alg1} finding each and every execution path in a program. Its source code is available in Appendix~\ref{app:code}. It tentatively disassembles one instruction at every program byte, and checks whether it yields a valid instruction. It then inserts these valid instructions into a graph, whose nodes are defined by their addresses and the outgoing edges by the values that the program counter might take after the execution of the instruction. For example, conditional jumps may have two outgoing edges, while data processing instructions may only have one outgoing edge to the immediately following instruction in the program. 

\begin{figure}[!ht]
    \centering
    \vspace*{-0.8cm}
    \includegraphics[width=0.95\textwidth]{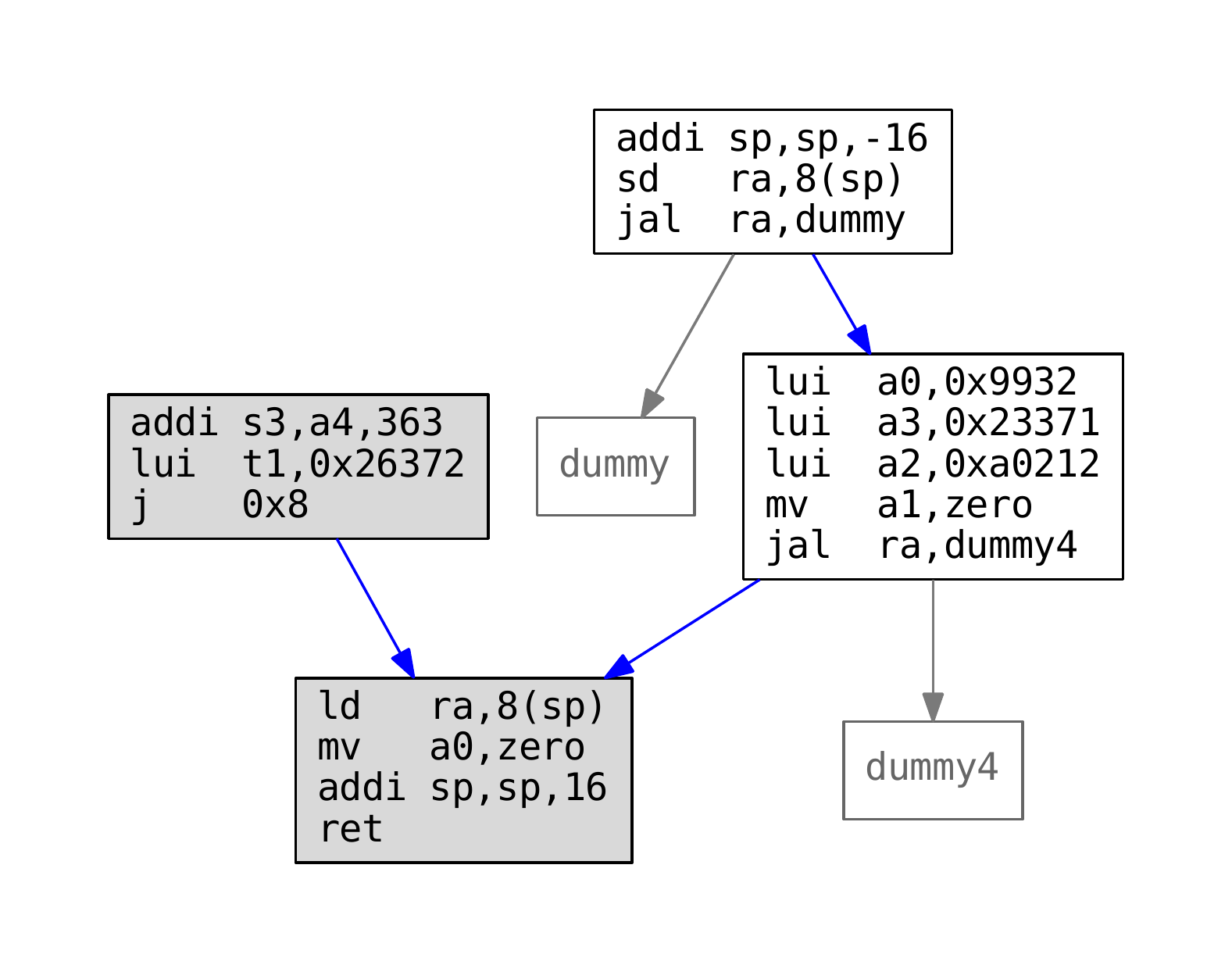}
    \vspace*{-0.8cm}
    \caption{The \texttt{function15c} (first presented in Fig.~\ref{fig:function1}) as shown by our disassembler. Unnecessary details such as instruction addresses or hexadecimal representations have been deleted. The gadget is highlighted in gray, and the dummy functions are shown in light-gray.}
    \vspace*{-0.5cm}
    \label{fig:disasm}
\end{figure}

Indirect jumps (like \texttt{ret}) do not have outgoing edges as the value of the program counter may not be known statically. We mark such instructions as \emph{Points of Interest} (or PoI, term coined in \cite{IRGadget}), to keep only the instructions that can reach one of those PoIs. Indeed, instruction sequences may only either reach a PoI, loop indefinitely or trigger an invalid instruction causing the program to crash. This can equivalently be rephrased as keeping only the subgraph coreachable from those PoIs. Additional work can be performed on this graph, like merging chains of nodes, yielding a \emph{control-flow graph} (CFG) showing both the MEP and HEP. We show in Fig.~\ref{fig:disasm} an example of such CFG. 

We used this algorithm to find such gadgets in the C standard library. Out of the 1957 unaligned sequences ending with a fixed jump offset, only one can realistically be used as a gadget in a traditional ROP attack. The scarcity of such gadgets on RISC-V architecture confirms our need for magic constants when encoding the gadgets in Section~\ref{sec:backdoor}. Indeed the opcode of a HEP instruction lies in the operand of the MEP instruction. 

Some more radical solutions would consist in trying to prevent overlapping code in RISC-V, either by deleting the compressed instruction C extension, or by requiring 32-bit instructions to be naturally aligned, or by changing the ISA so that the length of the instruction is encoded in first bit of every half-word. Though, we may lose one bit per half-word, hampering with the range of opcodes, \textit{i.e.} less immediates, or less registers. Furthermore, this requires extensive changes to the instruction set, and we believe that such a solution could only be implemented on next generation ISAs. 

\begin{algorithm}[!ht]
		\KwIn{$B_0, ... B_n$, a binary program}
		\KwResult{$G$, a directed graph of all execution paths}
		$G \stackrel{\mbox{\scriptsize{def}}}{=} (V,E)$\;
		$End \stackrel{\mbox{\scriptsize{def}}}{=} \emptyset$\;
		\For{$pc \stackrel{\mbox{\scriptsize{def}}}{=} 0$ \textnormal{\textbf{to}} $n$}
		{
			$I$ := Disasm\_one\_inst($B_{pc}, ...$)\;
			\If{\textnormal{$I$ \textbf{is not} a valid instruction}}
			{
				\textnormal{\textbf{continue}}
			}
			$V$.insert($pc$)\;
			\ForEach{$pc'$ \textnormal{\textbf{in}} $I$\textnormal{.get\_next\_pc()}}
			{
			    $E$.insert($pc$, $pc'$)
			}
			\If{\textnormal{$I$ \textbf{is} an indirect jump} }
			{
			    $End$.insert($pc$)
			}
			
		}
		$G' \stackrel{\mbox{\scriptsize{def}}}{=}$ coreachable($G$, $End$) \;
		\textnormal{\textbf{return}} $G'$\;
		\caption{Disassembly algorithm finding all execution paths in a binary.}
		\label{alg1}
	\end{algorithm}
	
\section{Related Work}
Andriesse et al.~\cite{Andriesse} have shown a method to hide malicious code using overlapping instructions in x86. It splits the code into smaller fragments and bruteforces a prefix and a suffix, for which the code fragment becomes a valid x86 MEP. This bruteforce method relies on the high density of the x86 instruction set, although it still sometimes requires manual intervention to conceal the fragments. The resulting hidden fragments are only one LCSAJ long, and always end by an indirect jump, hence easily caught by any ROP gadget searcher. Our approach allows better stealth by splitting the hidden code over several LCSAJs, for which the bruteforce method may not work anymore. We also apply our method to a RISC architecture, which does not benefit from the same code density.

ROP attacks have been subject to many academic studies since their first publication in 2007 \cite{Shacham} introducing the Galileo algorithm. Many variants based on the same algorithm have been published, like gadgets ending with indirect jumps \cite{JOP2011}, gadgets popping signal-contexts from the stack instead of call-frames \cite{SROP}, or attacks using format string vulnerabilities \cite{SOP2013}. Amongst popular ROP gadget searchers, only two have added support for RISC-V - \emph{xrop} and \emph{Radare2}~\cite{XROP,Radare2}, both of them implementing the Galileo algorithm, falling short of detecting this new class of gadgets. The closest to our work could be \emph{ROPgadget}~\cite{ROPgadget}, which tentatively disassembles a fixed number of instructions starting from each byte of the program. This method is particularly inefficient compared to our algorithm and to Galileo, but it could find some gadgets spanning over several LCSAJs, if they are shorter than a given threshold (by default 10 instructions). Quite surprisingly, after finding them, ROPGadget discards those gadgets by default, unless passed the option \texttt{-{}-multibr}. The algorithm that we provided comprehensively solves this aspect of gadget detection by revealing any gadget, whatever their length or number of LCSAJs is.

More recently, Borrello et al. \cite{ROPNeedle} published a method to insert backdoors in programs with encrypted ROP gadgets and a small decryption procedure. While encryption methods provide a definitive proof that the malicious behavior will indeed be hidden to static analysis, this does not address the problem of detection, as the decryption procedure is not concealed, and thus may be detected by static analysis. In this paper, we provided another method for adding such backdoors, without having any unconcealed element in the program. To achieve this result, we rely on a fine understanding of how current detectors work, exploiting their inability to find gadgets spanning over more than one LCSAJ.

\section{Conclusion and Future Work}

ROP attacks still pose a threat, although despite the wide deployment of dedicated countermeasures. Those protections fail to provide a satisfactory solution to these attacks, as we managed to design a new type of gadget on RISC-V, undetectable by existing tools, made of several linear-code sequences and jumps, that bypasses ASLR, DEP, stack canaries, G-Free and some compiler-based backward-edge CFI. We showed how to use such gadgets in two different attack schemes concealing a backdoor to perform privilege escalation attack on two standard Linux operating systems. Although the gadgets are written in C, we believe that it can generalize to other languages, such as JIT compilers once they become available on RISC-V, as well as other architectures featuring code overlap.

We provided a new algorithm aiming to replace previous Galileo based algorithms, that manages to find all the hidden execution paths of a program, and not just the last LCSAJ. This algorithm may be used both for offensive and defensive purposes.  However, we believe that its defensive usage is only provisional, as a definitive solution to prevent code overlap requires thorough changes in the ISA, which may only be implemented on next-generation architectures. 

\bibliographystyle{plain}
\bibliography{paper}

\vspace{-0.8cm}
\appendix
\section{Source code and artifact}
\label{app:code}
The C source code used for generating the gadgets, as well as links to the images of the Fedora and Debian virtual machines are available on the following link:
{\small\url{https://github.com/GAJaloyan/asiaccs2020}}.

\end{document}

%% file: customlibs.tex
\pgfdeclarepatternformonly{south west lines}{\pgfqpoint{-0pt}{-0pt}}{\pgfqpoint{3pt}{3pt}}{\pgfqpoint{3pt}{3pt}}{
    \pgfsetlinewidth{0.4pt}
    \pgfpathmoveto{\pgfqpoint{0pt}{0pt}}
    \pgfpathlineto{\pgfqpoint{3pt}{3pt}}
    \pgfpathmoveto{\pgfqpoint{2.8pt}{-.2pt}}
    \pgfpathlineto{\pgfqpoint{3.2pt}{.2pt}}
    \pgfpathmoveto{\pgfqpoint{-.2pt}{2.8pt}}
    \pgfpathlineto{\pgfqpoint{.2pt}{3.2pt}}
    \pgfusepath{stroke}}
        
\lstdefinelanguage
[rv64gc]{Assembler}
{alsoletter={.},
	morekeywords=
	{sra,srli,mv,xor,sw,sd,ret,nop, slti, jmp,sub,
	addi,addiw,c.li,c.lui,fld,ld,li,lui,lw,amoand.d,amoand.d.aq,amoor.w.aq,
	amoand.d.aqrl,amoand.d.rl,amoand.w,amoand.w.aq,amoand.w.aqrl,
	amoand.w.rl,amoor.d,amoor.d.aq,amoor.d.aqrl,amoor.d.rl,amoor.w,
	amoor.w.aq,amoor.w.aqrl,amoor.w.rl,bgtz,ble,bleu,blez,blt,bltu,
	csrc,csrci,csrr,csrrc,csrrci,csrrs,csrrsi,csrrwi,csrwi,fcvt.d.q,
	fmadd.d,fmadd.q,fmadd.s,fmsub.d,fmsub.q,fmsub.s,fnmadd.d,fnmadd.q,
	fnmadd.s,fnmsub.d,fnmsub.q,fnmsub.s,j,jal,lui,sra,ecall,mul,sltiu},morecomment=[l]{//},	morecomment=[s]{/*}{*/}}

\makeatletter
\g@addto@macro{\UrlBreaks}{\UrlOrds} 
\makeatother
\mathchardef\UrlBreakPenalty=100
\mathchardef\UrlBigBreakPenalty=1000
\newcommand\EatDot[1]{}

\makeatletter
\pgfarrowsdeclare{latexnew}{latexnew}
{
	\ifdim\pgfgetarrowoptions{latexnew}=-1pt%
	\pgfutil@tempdima=0.28pt%
	\pgfutil@tempdimb=\pgflinewidth%
	\ifdim\pgfinnerlinewidth>0pt%
	\pgfmathsetlength\pgfutil@tempdimb{.6\pgflinewidth-.4*\pgfinnerlinewidth}%
	\fi%
	\advance\pgfutil@tempdima by.3\pgfutil@tempdimb%
	\else%
	\pgfutil@tempdima=\pgfgetarrowoptions{latexnew}%
	\divide\pgfutil@tempdima by 10%
	\fi%
	\pgfarrowsleftextend{+-1\pgfutil@tempdima}%
	\pgfarrowsrightextend{+9\pgfutil@tempdima}%
}
{
	\ifdim\pgfgetarrowoptions{latexnew}=-1pt%
	\pgfutil@tempdima=0.28pt%
	\pgfutil@tempdimb=\pgflinewidth%
	\ifdim\pgfinnerlinewidth>0pt%
	\pgfmathsetlength\pgfutil@tempdimb{.6\pgflinewidth-.4*\pgfinnerlinewidth}%
	\fi%
	\advance\pgfutil@tempdima by.3\pgfutil@tempdimb%
	\else%
	\pgfutil@tempdima=\pgfgetarrowoptions{latexnew}%
	\divide\pgfutil@tempdima by 10%
	\pgfsetlinewidth{0bp}%
	\fi%
	\pgfpathmoveto{\pgfqpoint{9\pgfutil@tempdima}{0pt}}
	\pgfpathcurveto
	{\pgfqpoint{6.3333\pgfutil@tempdima}{.5\pgfutil@tempdima}}
	{\pgfqpoint{2\pgfutil@tempdima}{2\pgfutil@tempdima}}
	{\pgfqpoint{-1\pgfutil@tempdima}{3.75\pgfutil@tempdima}}
	\pgfpathlineto{\pgfqpoint{-1\pgfutil@tempdima}{-3.75\pgfutil@tempdima}}
	\pgfpathcurveto
	{\pgfqpoint{2\pgfutil@tempdima}{-2\pgfutil@tempdima}}
	{\pgfqpoint{6.3333\pgfutil@tempdima}{-.5\pgfutil@tempdima}}
	{\pgfqpoint{9\pgfutil@tempdima}{0pt}}
	\pgfusepathqfill
}

\pgfarrowsdeclarereversed{latexnew reversed}{latexnew reversed}{latexnew}{latexnew}

\pgfarrowsdeclare{latex'new}{latex'new}
{
	\ifdim\pgfgetarrowoptions{latex'new}=-1pt%
	\pgfutil@tempdima=0.28pt%
	\advance\pgfutil@tempdima by.3\pgflinewidth%
	\else%
	\pgfutil@tempdima=\pgfgetarrowoptions{latex'new}%
	\divide\pgfutil@tempdima by 10%
	\fi%
	\pgfarrowsleftextend{+-4\pgfutil@tempdima}
	\pgfarrowsrightextend{+6\pgfutil@tempdima}
}
{
	\ifdim\pgfgetarrowoptions{latex'new}=-1pt%
	\pgfutil@tempdima=0.28pt%
	\advance\pgfutil@tempdima by.3\pgflinewidth%
	\else%
	\pgfutil@tempdima=\pgfgetarrowoptions{latex'new}%
	\divide\pgfutil@tempdima by 10%
	\pgfsetlinewidth{0bp}%
	\fi%
	\pgfpathmoveto{\pgfqpoint{6\pgfutil@tempdima}{0\pgfutil@tempdima}}
	\pgfpathcurveto
	{\pgfqpoint{3.5\pgfutil@tempdima}{.5\pgfutil@tempdima}}
	{\pgfqpoint{-1\pgfutil@tempdima}{1.5\pgfutil@tempdima}}
	{\pgfqpoint{-4\pgfutil@tempdima}{3.75\pgfutil@tempdima}}
	\pgfpathcurveto
	{\pgfqpoint{-1.5\pgfutil@tempdima}{1\pgfutil@tempdima}}
	{\pgfqpoint{-1.5\pgfutil@tempdima}{-1\pgfutil@tempdima}}
	{\pgfqpoint{-4\pgfutil@tempdima}{-3.75\pgfutil@tempdima}}
	\pgfpathcurveto
	{\pgfqpoint{-1\pgfutil@tempdima}{-1.5\pgfutil@tempdima}}
	{\pgfqpoint{3.5\pgfutil@tempdima}{-.5\pgfutil@tempdima}}
	{\pgfqpoint{6\pgfutil@tempdima}{0\pgfutil@tempdima}}
	\pgfusepathqfill
}

\pgfarrowsdeclarereversed{latex'new reversed}{latex'new reversed}{latex'new}{latex'new}

\pgfarrowsdeclare{onew}{onew}
{
	\pgfarrowsleftextend{+-.5\pgflinewidth}
	\ifdim\pgfgetarrowoptions{onew}=-1pt%
	\pgfutil@tempdima=0.4pt%
	\advance\pgfutil@tempdima by.2\pgflinewidth%
	\pgfutil@tempdimb=9\pgfutil@tempdima\advance\pgfutil@tempdimb by.5\pgflinewidth%
	\pgfarrowsrightextend{+\pgfutil@tempdimb}%
	\else%
	\pgfutil@tempdima=\pgfgetarrowoptions{onew}%
	\advance\pgfutil@tempdima by -0.5\pgflinewidth%
	\pgfarrowsrightextend{+\pgfutil@tempdima}%
	\fi%
}
{ 
	\ifdim\pgfgetarrowoptions{onew}=-1pt%
	\pgfutil@tempdima=0.4pt%
	\advance\pgfutil@tempdima by.2\pgflinewidth%
	\pgfutil@tempdimb=0pt%
	\else%
	\pgfutil@tempdima=\pgfgetarrowoptions{onew}%
	\divide\pgfutil@tempdima by 9%
	\pgfutil@tempdimb=0.5\pgflinewidth%
	\fi%
	\pgfsetdash{}{+0pt}
	\pgfpathcircle{\pgfpointadd{\pgfqpoint{4.5\pgfutil@tempdima}{0bp}}%
		{\pgfqpoint{-\pgfutil@tempdimb}{0bp}}}%
	{4.5\pgfutil@tempdima-\pgfutil@tempdimb}%
	\pgfusepathqstroke
}

\pgfarrowsdeclare{squarenew}{squarenew}
{
	\ifdim\pgfgetarrowoptions{squarenew}=-1pt%
	\pgfutil@tempdima=0.4pt
	\advance\pgfutil@tempdima by.275\pgflinewidth%
	\pgfarrowsleftextend{+-\pgfutil@tempdima}
	\advance\pgfutil@tempdima by.5\pgflinewidth
	\pgfarrowsrightextend{+\pgfutil@tempdima}
	\else%
	\pgfutil@tempdima=\pgfgetarrowoptions{squarenew}%
	\divide\pgfutil@tempdima by 8%
	\pgfarrowsleftextend{+-7\pgfutil@tempdima}%
	\pgfarrowsrightextend{+1\pgfutil@tempdima}%
	\fi%
}
{
	\ifdim\pgfgetarrowoptions{squarenew}=-1pt%
	\pgfutil@tempdima=0.4pt%
	\advance\pgfutil@tempdima by.275\pgflinewidth%
	\pgfutil@tempdimb=0pt%
	\else%
	\pgfutil@tempdima=\pgfgetarrowoptions{squarenew}%
	\divide\pgfutil@tempdima by 8%
	\pgfutil@tempdimb=0.5\pgflinewidth%
	\fi%
	\pgfsetdash{}{+0pt}
	\pgfsetroundjoin
	\pgfpathmoveto{\pgfpointadd{\pgfqpoint{1\pgfutil@tempdima}{4\pgfutil@tempdima}}
		{\pgfqpoint{-\pgfutil@tempdimb}{-\pgfutil@tempdimb}}}
	\pgfpathlineto{\pgfpointadd{\pgfqpoint{-7\pgfutil@tempdima}{4\pgfutil@tempdima}}
		{\pgfqpoint{\pgfutil@tempdimb}{-\pgfutil@tempdimb}}}
	\pgfpathlineto{\pgfpointadd{\pgfqpoint{-7\pgfutil@tempdima}{-4\pgfutil@tempdima}}
		{\pgfqpoint{\pgfutil@tempdimb}{\pgfutil@tempdimb}}}
	\pgfpathlineto{\pgfpointadd{\pgfqpoint{1\pgfutil@tempdima}{-4\pgfutil@tempdima}}
		{\pgfqpoint{-\pgfutil@tempdimb}{\pgfutil@tempdimb}}}
	\pgfpathclose
	\pgfusepathqfillstroke
}

\pgfarrowsdeclare{stealthnew}{stealthnew}
{
	\ifdim\pgfgetarrowoptions{stealthnew}=-1pt%
	\pgfutil@tempdima=0.28pt%
	\pgfutil@tempdimb=\pgflinewidth%
	\ifdim\pgfinnerlinewidth>0pt%
	\pgfmathsetlength\pgfutil@tempdimb{.6\pgflinewidth-.4*\pgfinnerlinewidth}%
	\fi%
	\advance\pgfutil@tempdima by.3\pgfutil@tempdimb%
	\else%
	\pgfutil@tempdima=\pgfgetarrowoptions{stealthnew}%
	\divide\pgfutil@tempdima by 8%
	\fi%
	\pgfarrowsleftextend{+-3\pgfutil@tempdima}
	\pgfarrowsrightextend{+5\pgfutil@tempdima}
}
{
	\ifdim\pgfgetarrowoptions{stealthnew}=-1pt%
	\pgfutil@tempdima=0.28pt%
	\pgfutil@tempdimb=\pgflinewidth%
	\ifdim\pgfinnerlinewidth>0pt%
	\pgfmathsetlength\pgfutil@tempdimb{.6\pgflinewidth-.4*\pgfinnerlinewidth}%
	\fi%
	\advance\pgfutil@tempdima by.3\pgfutil@tempdimb%
	\else%
	\pgfutil@tempdima=\pgfgetarrowoptions{stealthnew}%
	\divide\pgfutil@tempdima by 8%
	\pgfsetlinewidth{0bp}%
	\fi%
	\pgfpathmoveto{\pgfqpoint{5\pgfutil@tempdima}{0pt}}
	\pgfpathlineto{\pgfqpoint{-3\pgfutil@tempdima}{4\pgfutil@tempdima}}
	\pgfpathlineto{\pgfpointorigin}
	\pgfpathlineto{\pgfqpoint{-3\pgfutil@tempdima}{-4\pgfutil@tempdima}}
	\pgfusepathqfill
}

\pgfarrowsdeclarereversed{stealthnew reversed}{stealthnew reversed}{stealthnew}{stealthnew}

\pgfarrowsdeclare{tonew}{tonew}
{
	\ifdim\pgfgetarrowoptions{tonew}=-1pt%
	\pgfutil@tempdima=0.84pt%
	\advance\pgfutil@tempdima by1.3\pgflinewidth%
	\pgfutil@tempdimb=0.21pt%
	\advance\pgfutil@tempdimb by.625\pgflinewidth%
	\else%
	\pgfutil@tempdima=\pgfgetarrowoptions{tonew}%
	\pgfarrowsleftextend{+-0.8\pgfutil@tempdima}%
	\pgfarrowsrightextend{+0.2\pgfutil@tempdima}%
	\fi%
}
{
	\ifdim\pgfgetarrowoptions{tonew}=-1pt%
	\pgfutil@tempdima=0.28pt%
	\advance\pgfutil@tempdima by.3\pgflinewidth%
	\pgfutil@tempdimb=0pt,%
	\else%
	\pgfutil@tempdima=\pgfgetarrowoptions{tonew}%
	\multiply\pgfutil@tempdima by 100%
	\divide\pgfutil@tempdima by 375%
	\pgfutil@tempdimb=0.4\pgflinewidth%
	\fi%
	\pgfsetdash{}{+0pt}
	\pgfsetroundcap
	\pgfsetroundjoin
	\pgfpathmoveto{\pgfpointorigin}
	\pgflineto{\pgfpointadd{\pgfpoint{0.75\pgfutil@tempdima}{0bp}}
		{\pgfqpoint{-2\pgfutil@tempdimb}{0bp}}}
	\pgfusepathqstroke
	\pgfsetlinewidth{0.8\pgflinewidth}
	\pgfpathmoveto{\pgfpointadd{\pgfqpoint{-3\pgfutil@tempdima}{4\pgfutil@tempdima}}
		{\pgfqpoint{\pgfutil@tempdimb}{0bp}}}
	\pgfpathcurveto
	{\pgfpointadd{\pgfqpoint{-2.75\pgfutil@tempdima}{2.5\pgfutil@tempdima}}
		{\pgfqpoint{0.5\pgfutil@tempdimb}{0bp}}}
	{\pgfpointadd{\pgfqpoint{0pt}{0.25\pgfutil@tempdima}}
		{\pgfqpoint{-0.5\pgfutil@tempdimb}{0bp}}}
	{\pgfpointadd{\pgfqpoint{0.75\pgfutil@tempdima}{0pt}}
		{\pgfqpoint{-\pgfutil@tempdimb}{0bp}}}
	\pgfpathcurveto
	{\pgfpointadd{\pgfqpoint{0pt}{-0.25\pgfutil@tempdima}}
		{\pgfqpoint{-0.5\pgfutil@tempdimb}{0bp}}}
	{\pgfpointadd{\pgfqpoint{-2.75\pgfutil@tempdima}{-2.5\pgfutil@tempdima}}
		{\pgfqpoint{0.5\pgfutil@tempdimb}{0bp}}}
	{\pgfpointadd{\pgfqpoint{-3\pgfutil@tempdima}{-4\pgfutil@tempdima}}
		{\pgfqpoint{\pgfutil@tempdimb}{0bp}}}
	\pgfusepathqstroke
}

\pgfarrowsdeclarealias{<new}{>new}{tonew}{tonew}

\makeatother

\pgfsetarrowoptions{latexnew}{-1pt}
\pgfsetarrowoptions{latex'new}{-1pt}
\pgfsetarrowoptions{onew}{-1pt}
\pgfsetarrowoptions{squarenew}{-1pt}
\pgfsetarrowoptions{stealthnew}{-1pt}
\pgfsetarrowoptions{tonew}{-1pt}
\pgfkeys{/tikz/.cd, arrowhead/.default=-1pt, arrowhead/.code={
		\pgfsetarrowoptions{latexnew}{#1},
		\pgfsetarrowoptions{latex'new}{#1},
		\pgfsetarrowoptions{onew}{#1},
		\pgfsetarrowoptions{squarenew}{#1},
		\pgfsetarrowoptions{stealthnew}{#1},
		\pgfsetarrowoptions{tonew}{#1},
}}

%% file: ropfig.tex
\colorlet{myhl}{blue!20}
\begin{tikzpicture}[scale=1.0, every node/.style={scale=1}]
	
	\fill [myhl] (1.9,-1.5) rectangle (3,-2);
	\fill [myhl] (0,-2) rectangle (3,-10);
	
	\draw (0,0) rectangle (3,-11);
	
	\draw[decoration={brace,raise=1pt, mirror,amplitude=10pt},decorate,line width=1pt] 
	(0,0) -- (0,-4) node [black, midway, xshift=-1.3cm] {\small stack frame \phantom{a}} ;
	\draw[decoration={brace,raise=1pt, mirror,amplitude=10pt},decorate,line width=1pt,blue] 
	(0,-4) -- (0,-5.5) node [blue, midway, xshift=-1.3cm] {\small stack frame \phantom{a}} ;
	\draw[decoration={brace,raise=1pt, mirror,amplitude=10pt},decorate,line width=1pt,blue] 
	(0,-5.5) -- (0,-7) node [blue, midway, xshift=-1.3cm] {\small stack frame \phantom{a}} ;
	\draw[decoration={brace,raise=1pt, mirror,amplitude=10pt},decorate,line width=1pt,blue] 
	(0,-7) -- (0,-10) node [blue, midway, xshift=-1.3cm] {\small stack frame \phantom{a}} ;

	\node [gray] at (1.5,-1.75) {\tiny \phantom{padding padding} padding};
	\node [gray] at (1.5,-2.25) {\tiny padding padding padding};
	\node at (1.5,-1.25) {local data};
	\draw (0,-2.5) to (3,-2.5);
	\node [text width=3.3cm,text centered] at (1.5,-3.25) { {\small \centering return address\\\vspace*{-0.1cm} \textcolor{black!60}{gadget\_1}}};
	\draw (0,-4) to (3,-4);
	\node [text width=3.3cm,text centered] at (1.5,-4.75) { {\small \centering return address\\\vspace*{-0.1cm} \textcolor{black!60}{gadget\_2}}};
	\draw (0,-5.5) to (3,-5.5);
	\node [text width=3.3cm,text centered] at (1.5,-6.25) { {\small \centering return address\\\vspace*{-0.1cm} \textcolor{black!60}{gadget\_3}}};
	\draw (0,-7) to (3,-7);
	\node [text width=3.3cm,text centered] at (1.5,-7.75) { {\small \centering local data\\\vspace*{-0.0cm} \textcolor{black!60}{"/bin/sh"}}};
	\draw (0,-8.5) to (3,-8.5);
	\node [text width=3.3cm,text centered] at (1.5,-9.25) { {\small \centering return address\\\vspace*{-0.1cm} \textcolor{black!60}{system\_call}}};
	\draw (0,-10) to (3,-10);

	\node at (1.5,-10.5) {...};	
	
	\draw (5,0) rectangle (7.8,-2);
	\node[below right] at (5,0) {\small\begin{lstlisting}[linewidth=6cm]
...
jal strcpy
...
ret
\end{lstlisting}};
	
	\draw (5,-2.75) rectangle (7.8,-4.75);
	\node [above] at (6.4,-2.85) {\footnotesize gadget\_1};
	\node[below right] at (5,-2.8) {\small\begin{lstlisting}
mv  a1,0
mv  a2,0
...
ret
\end{lstlisting}};
	
	\draw (5,-5.5) rectangle (7.8,-7);
	\node [above] at (6.4,-5.6) {\footnotesize gadget\_2};
	\node[below right] at (5,-5.5) {\small\begin{lstlisting}
mv  a7,221
...
ret
\end{lstlisting}};

	\draw (5,-7.75) rectangle (7.8,-9.25);
	\node [above] at (6.4,-7.85) {\footnotesize gadget\_3};
	\node[below right] at (5,-7.75) {\small\begin{lstlisting}
mv  a0,sp
...
ret
\end{lstlisting}};
	
	\draw (5,-10) rectangle (7.8,-10.5);
	\node [above] at (6.4,-10.1) {\footnotesize system\_call};
	\node[below right] at (5,-9.9) {\small\begin{lstlisting}
ecall
\end{lstlisting}};
	
	\draw [-latexnew,line width=1pt, arrowhead=0.2cm,dotted] (5,-1.8) -- (3,-3.15);
	\draw [-latexnew,line width=1pt, arrowhead=0.2cm] (3,-3.35) -- (5,-3.15);
	\draw [-latexnew,line width=1pt, arrowhead=0.2cm,dotted] (5,-4.4) -- (3,-4.65);
	\draw [-latexnew,line width=1pt, arrowhead=0.2cm] (3,-4.85) -- (5,-5.85);
	\draw [-latexnew,line width=1pt, arrowhead=0.2cm,dotted] (5,-6.8) -- (3,-6.15);
	\draw [-latexnew,line width=1pt, arrowhead=0.2cm] (3,-6.35) -- (5,-8.1);
	\draw [-latexnew,line width=1pt, arrowhead=0.2cm,dotted] (5,-9) -- (3,-9.15);
	\draw [-latexnew,line width=1pt, arrowhead=0.2cm] (3,-9.35) -- (5,-10.3);

	\draw [-latexnew,line width=2pt, arrowhead=0.25cm, blue] (5.1,-0.9) to [bend right] node [midway,text centered,above,blue] {\centering \small \phantom{aa}overflow}  (2.5,-1.5);
	
\end{tikzpicture}

%% file: gadget.tex
\definecolor{mygrey}{RGB}{210,210,210}

\begin{tikzpicture}
    \node at (0, 0.5) {\textsc{Function15c}};
    \node at (-3, 0.5) {\textsc{MEP}};
    \node at (3, 0.5) {\textsc{HEP}};
    
\fill [mygrey] (1.5,-1.5) rectangle (6,-3);
\fill [mygrey] (-6,-4) rectangle (-1.5,-6);
\draw[pattern=south west lines, pattern color=black!30!white] (1.5,0) rectangle (6,-1.5);
\draw[pattern=south west lines, pattern color=black!30!white] (1.5,-3) rectangle (6,-6);

    \draw (-6,0) rectangle (6, -6);
    \draw (-1.5,0) to (-1.5,-6);
    \draw (1.5,0) to (1.5,-6);
    \draw (-6,-1) to (6, -1);
    \draw (-6,-1.5) to (6, -1.5);
    \draw (-6,-3) to (6, -3);
    \draw (-6,-4) to (6, -4);
    
    \node at (0,-0.5) {save sequence};
    \node[below right] at (-6,0.1) {\begin{lstlisting}
addi    sp,sp,-16
sd      ra,8(sp)
\end{lstlisting}};

    \node at (0,-1.275) {dummy call};
    \node[below right] at (-6,-0.9) {\begin{lstlisting}
jal     ra,dummy
\end{lstlisting}};

    \node at (0,-2.25) {overlapping code};
    \node[below right] at (-6,-1.4) {\begin{lstlisting}
lui     a0,0x9932
lui     a3,0x23371
lui     a2,0xa0212
\end{lstlisting}};
    \node[below right] at (1.5,-1.4) {\begin{lstlisting}
addi    s3,a4,363
lui     t1,0x26372
jmp     0x8
\end{lstlisting}};
    \node at (0,-3.5) {instructions};
    \node[below right] at (-6,-2.88) {\begin{lstlisting}
mv      a1,zero
jal     ra,dummy4
\end{lstlisting}};

    \node at (0,-4.9) {restore sequence};
    \node[below right] at (-6,-3.92) {\begin{lstlisting}
ld      ra,8(sp)
mv      a0,zero
addi    sp,sp,16
ret
\end{lstlisting}};

\end{tikzpicture}